 \documentclass[]{elsart}

\usepackage[authoryear,round]{natbib}
\usepackage{amssymb,bm}
\usepackage{graphicx}

\begin{document}

\begin{frontmatter}



\title{ Analytical calculation of Coulomb corrections to $e^+e^-$ pair production
at intermediate photon energies}

\author{R.N. Lee},
\ead{R.N.Lee@inp.nsk.su}
\author{A. I. Milstein},
\ead{A.I.Milstein@inp.nsk.su}
and \author{V.M. Strakhovenko}
\ead{V.M.Strakhovenko@inp.nsk.su}
\address{Budker Institute of Nuclear Physics, 630090 Novosibirsk, Russia}

\begin{abstract}
First correction to the high-energy asymptotics of the total $e^+e^-$
photoproduction cross section in the electric field of a heavy atom is obtained
with the exact account of this field. The consideration is based on the use of
the quasiclassical electron Green function in an external electric field. The
influence of screening on the Coulomb corrections is examined in the leading
approximation. It turns out that the high-energy asymptotics of the
corresponding correction is independent of the photon energy. The detailed
comparison of our results with experimental data is performed. This comparison
has justified the analytical result and allowed us to elaborate a simple ansatz
for the next-to-leading correction. Using this ansatz, good agreement with the
experimental data is obtained for photon energies above a few $MeV$. In the
region where both produced particles are relativistic, the corrections to the
high-energy asymptotics of the electron (positron) spectrum are obtained. In
addition, analogous corrections to the bremsstrahlung spectrum are derived
starting from the corresponding results for pair production.
\end{abstract}



\end{frontmatter}

\section{Introduction}

Knowledge of the photoabsorption cross sections is very important in various
applications, see, e.g., \citep{Hubbell2000}. The relevant processes are the
atomic photoeffect, nuclear photoabsorption, incoherent and coherent photon
scattering and $e^+e^-$ pair production. In the coherent processes, by
definition, there is no excitation or ionization of an atom. The high-accuracy
estimation of the corresponding cross sections is required. They have different
dependence on the photon energy $\omega$. At $\omega\gtrsim 10 MeV$, the cross
section  of $e^+e^-$ pair production becomes dominant \citep{HGO1980}. The
coherent contribution $\sigma_{coh}$ to the pair production cross section is
roughly $Z$ times larger than the incoherent one ($Z$ is the atomic number),
thereby being the most important for heavy atoms. Just the coherent pair
production is considered below.

The theoretical and experimental investigation of the coherent pair production
has a long history, see \citep{HGO1980}. In the Born approximation, the cross
section $\sigma_B$ is known for arbitrary photon energy
\citep{BH1934,Racah1934}. The account of the effect of screening is
straightforward in this approximation and can be easily performed if the atomic
form factor is known \citep{JLS1950}. For heavy atoms it is necessary to take
into account the Coulomb corrections $\sigma_C$,
\begin{equation}
\sigma_{coh}=\sigma_B+\sigma_C\,.
\end{equation}

These corrections are higher order terms of the perturbation theory with
respect to the atomic field. The magnitude of $\sigma_C$ depends on $\omega$
and the parameter $Z\alpha$ ($\alpha=1/137$ is the fine-structure constant).
The formal expression for $\sigma_C$, exact in $Z\alpha$ and $\omega$, was
derived by \citet{Overbo1968}. This expression has a very complicated form
causing severe difficulties in computations. The difficulties grow as $\omega$
increases, so that numerical results in \citep{Overbo1968} were obtained only
for $\omega<5 MeV$.

For the high-energy region $\omega\gg m$ ($m$ is the electron mass), the
consideration is greatly simplified. As a result, a rather simple form was
obtained in \citep{BM1954,DBM1954} for the Coulomb corrections in the leading
approximation with respect to $m/\omega$. However, the theoretical description
of the Coulomb corrections at intermediate photon energies ($5\div100 MeV$) has
not been completed. At present, all estimates of $\sigma_C$ in this region are
based on the "bridging" expression derived by \citet{Overbo1977}. This
expression is actually an extrapolation of the results obtained for
$\omega<5MeV$. It is based on some assumptions on the form of the asymptotic
expansion of $\sigma_C$ at high photon energy. It is commonly believed that the
"bridging" expression has an accuracy providing the maximum error in
$\sigma_{coh}$ of the order of a few tens of percent.

Here we develop a description of $e^+e^-$-pair production at intermediate
photon energies by deriving the next-to-leading term of the high-energy
expansion of $\sigma_C$. First we consider a pure Coulomb field and represent
$\sigma_C$ in the form
\begin{equation}\label{eq:expansion}
\sigma_C=\sigma_C^{(0)}+\sigma_C^{(1)}+\sigma_C^{(2)}+\ldots
\end{equation}
The term $\sigma_C^{(n)}$ has the form $(m/\omega)^n S^{(n)}(\ln \omega/m)$,
where $S^{(n)}(x)$ is some polynomial. The $\omega$-independent term
$\sigma_C^{(0)}$ corresponds to the result of \citet{DBM1954}. In the present
paper we derive the term $\sigma_C^{(1)}$. It turns out that $S^{(1)}$ is
$\omega$-independent in contrast to a second-degree polynomial suggested by
\citet{Overbo1977}. We propose a new ansatz for $\sigma_C^{(2)}$, which
provides a good agreement with available experimental data for $\omega>5 MeV$.

The high-energy expansion of the Coulomb corrections to the spectrum has the
same form as (\ref{eq:expansion}). In the region $\varepsilon_{\pm}\gg m$, we
derive the term $d\sigma_C^{(1)}/dx$, where $\varepsilon_-$ and $\varepsilon_+$
are the electron and positron energy, respectively, $x=\varepsilon_{-}/\omega$.
The term $d\sigma_C^{(1)}/dx$ may turn important, e.g., for description of the
development of electromagnetic showers in a medium. The correction found is
antisymmetric with respect to the permutation
$\varepsilon_{+}\leftrightarrow\varepsilon_{-}$ and does not contribute to the
total cross section. In fact,  $\sigma_C^{(1)}$ originates from two energy
regions $\varepsilon_{+}\sim m$ and $\varepsilon_{-}\sim m$, where the spectrum
is not known. However, we emphasize that it differs drastically from the result
obtained by \citet{DBM1954} for $\varepsilon_{\pm}\gg m$, if the latter is
formally applied at $\varepsilon_-\sim m$ or $\varepsilon_+\sim m$.

For the first time, we estimate the effect of screening on $\sigma_C$ at
$\omega\gg m$. In the leading approximation, we find the corresponding
correction $\sigma_C^{(scr)}$, which is $\omega$-independent similar to
$\sigma_C^{(0)}$. So, for the atomic field, $\sigma_C^{(scr)}$ should be added
to the right-hand side of Eq.  (\ref{eq:expansion}). The screening correction
to the spectrum is also obtained.

Tedious and cumbersome calculations have been performed to derive our formulas.
All technical details are omitted here and will be presented elsewhere.

\section{General discussion}

The cross section of $e^+e^-$ pair production by a photon in an
external field reads
\begin{equation}\label{eq:cs}
d\sigma_{coh}=\frac{\alpha}{(2\pi)^4\omega}\,d\bm{p}\,d\bm{q}\,\delta
(\omega - \varepsilon_+ -\varepsilon_-)|M|^{2}\,,
\end{equation}
where $\varepsilon_-=\varepsilon_{ p}=\sqrt{\bm p^2+m^2}$,
$\varepsilon_+=\varepsilon_{ q}$, and $\bm p$, $\bm q$ are the electron and
positron momenta, respectively. The matrix element $M$ has the form
\begin{equation}
M\,=\,\int d\bm r \,\bar \psi_{\bm p }^{(+)}(\bm r )\,\hat
{e}\,\psi _{\bm q}^{(-)}(\bm r )\exp{(i\bm k\bm r )}\,\,.
\end{equation}
Here $ \psi_{\bm p}^{(+)}$ and $\psi_{\bm q}^{(-)}$ are
positive-energy and negative-energy solutions of the Dirac
equation in the external field, $e_{\mu}$ is the photon
polarization 4-vector,   $\bm k$ is the photon
 momentum. It is
 convenient  to study various processes in  external
fields  using the Green function
$G(\bm{r}_2,\bm{r}_1|\varepsilon)$ of the Dirac equation in this
field. This Green function can be represented in the form
\begin{eqnarray}\label{eq:FG}
G(\bm r_2 ,\bm r_1|\,\varepsilon)\,&=&\,\sum_{n}
\frac{\psi_{n}^{(+)}(\bm r_2)\bar\psi_{n}^{(+)}(\bm r_1 )}
{\varepsilon -\varepsilon_{n}\, + i0} \nonumber\\
&&+ \int \frac{d\bm{p}}{(2\pi)^{3}}\,\left[\,\frac{\psi_{\bm p}^{(+)}(\bm r_2 )
\bar\psi_{\bm p}^{(+)}(\bm r_1 )}{\varepsilon -\varepsilon_{p}\,+ i0}\,+
\,\frac{\psi_{\bm p}^{(-)}(\bm r_2 )\bar\psi_{\bm p}^{(-)}(\bm r_1 )}
{\varepsilon +\varepsilon_{p}\,- i0}\,\right] \,\,,
\end{eqnarray}
where $\psi_{n}^{(+)}$ is the discrete-spectrum wave function,
$\varepsilon_{n}$ is the corresponding binding energy. The regularization of
denominators in (\ref{eq:FG}) corresponds to the Feynman rule. From
(\ref{eq:FG}),
\begin{eqnarray}\label{eq:FG1}
\int d\Omega_{\bm{q}}\ \psi_{\bm q}^{(-)}(\bm r_2 ) \bar\psi_{\bm q}^{(-)}(\bm
r_1 )&=& -i \frac{(2\pi)^2}{q\varepsilon_{ q}}
 \delta G\,(\bm r_2 ,\bm
r_1|-\varepsilon_{\bm q})\,,\nonumber\\
\int d\Omega_{\bm{p}}\ \psi_{\bm p}^{(+)}(\bm r_1 ) \bar\psi_{\bm
p}^{(+)}(\bm r_2 )&=& i \frac{(2\pi)^2}{p\varepsilon_{\bm p}}
 \delta G\,(\bm r_1 ,\bm r_2|\varepsilon_{ p})\,,
\end{eqnarray}
where $\Omega_{\bm{p}}$ is the solid angle of $\bm p$, and $\delta
G=G-\tilde{G}$. The function $\tilde{G}$ is obtained from
(\ref{eq:FG}) by the replacement $i0\leftrightarrow -i0$.

Taking the integrals over $\Omega_{\bm{p}}$ and $\Omega_{\bm{q}}$
in (\ref{eq:cs}), we obtain the electron  spectrum, which is
 the cross section differential with respect to the  electron energy
 $\varepsilon_-$ .  Using the relations
(\ref{eq:FG1}), we express this spectrum  via the Green functions:
\begin{eqnarray}\label{eq:se}
 \frac{d\sigma_{coh}}{d\varepsilon_-}=
 \frac{ \alpha}{\omega}
\int\!\!\int d\bm r_1\, d\bm r_2\,
\mbox{e}^{i\bm{kr}}\,Sp\,\left\{
 \delta G(\bm r_1,\bm r_2|\varepsilon_-)\,\hat{e}\,
 \delta G(\bm r_2,\bm
 r_1|-\varepsilon_+)\,\hat{e}^*\right\}\,,
\end{eqnarray}
where $\varepsilon_+=\omega-\varepsilon_-$ and $\bm r=\bm r_2-\bm
r_1$.

Due to the optical theorem, the process of pair production is
related to the process of Delbr\"uck scattering (coherent
scattering of a photon in the electric field of an atom via
virtual electron-positron pairs). At zero scattering angle, the
amplitude $M_D$ of Delbr\"uck scattering reads
\begin{eqnarray}\label{eq:Md}
M_D=2i\alpha \int\!\! d\varepsilon \int\!\!\!\!\int d\bm r_1\,
d\bm r_2\, \mbox{e}^{i\bm{kr}}\,Sp\,\left\{
 G(\bm r_1,\bm r_2|\varepsilon)\,\hat{e}\,
 G(\bm r_2,\bm r_1|\varepsilon-\omega)\,\hat{e}^*\right\}.
\end{eqnarray}

It follows from Eqs. (\ref{eq:se}),(\ref{eq:Md}) and the analytical properties
of the Green function that
\begin{equation}\label{eq:Opt}
\frac1\omega\mbox{Im}\,M_D=\sigma_{coh}+\sigma_{bf}\,.
\end{equation}
Here
\begin{eqnarray}\label{eq:bf}
&& \sigma_{bf}=
 -\frac{ 2i\pi\alpha}{\omega}
\int\!\!\int d\bm r_1\, d\bm r_2\, \mbox{e}^{i\bm{kr}}\sum_n\,
Sp\left\{ \rho_n(\bm r_1,\bm r_2)\,\hat{e}\,
 \delta G(\bm r_2,\bm
 r_1|\varepsilon_n-\omega)\,\hat{e}^*\right\},\nonumber\\
&&\rho_n(\bm r_1,\bm r_2)=\lim_{\varepsilon\to \varepsilon_n}
(\varepsilon- \varepsilon_n)G(\bm r_1,\bm
 r_2|\varepsilon).
\end{eqnarray}

The quantity $\sigma_{bf}$ coincides with the total cross section of the
so-called bound-free pair production when an electron is produced in a bound
state. In fact, due to the Pauli principle, there is no bound-free pair
production  on neutral  atoms. Nevertheless,  the term $\sigma_{bf}$ should be
kept   in the r.h.s. of (\ref{eq:Opt}).  In a Coulomb field, the total cross
section $\sigma_{bf}$  was obtained  in \citep{MS1993} for $\omega\gg m$. In
this limit, $\sigma_{bf}\propto 1/m\omega$ and should be taken into account
when using the relation (\ref{eq:Opt}) for the calculation of the corrections
to $\sigma_{coh}$ from. The main contribution to $\sigma_{bf}$ comes from the
low-lying bound states \citep{MS1993} when screening can be neglected. So, in
(\ref{eq:Opt}) we can use $\sigma_{bf}$ obtained in \citep{MS1993}.

\section{Coulomb corrections to the spectrum}
\label{sec:CCS}

In this section we consider the Coulomb corrections to the spectrum,
$d\sigma_{C}/dx$, for $\varepsilon_{\pm}\gg m$ taking into account terms of the
order $m/\varepsilon_{\pm}$. Within this accuracy, the spectrum is determined
by  the region of small angles between vectors $\bm p,\,\bm q$, and $\bm k$ in
(\ref{eq:cs}). In this case the angular momenta of both particles are large,
providing  the applicability of the  quasiclassical approximation. Thus we can
use the quasiclassical Green function in (\ref{eq:se}). For a pure Coulomb
field, this function was found in \citep{MS1983};  for arbitrary localized
field, it was obtained in \citep{LMS2000}. In the latter paper, the first
correction to the quasiclassical Green function was  derived as well.

According to \cite{DBM1954}, the higher order terms of the perturbation theory
with respect to the external field (Coulomb corrections) are not seriously
modified by screening. However, this question has not been studied
quantitatively so far. The influence of screening on Coulomb corrections is
investigated in detail in Section \ref{sec:Scr}. In the present Section we
calculate $d\sigma_C/d\varepsilon_-$ in a pure Coulomb field.

\begin{figure}
\centering\setlength{\unitlength}{0.1cm}
\begin{picture}(105,80)
 \put(56,0){\makebox(0,0)[t]{$x$}}
 \put(-6,30){\rotatebox[origin=c]{90}{$\sigma_0^{-1} d\sigma_C/dx$}}
\put(0,0){\includegraphics[width=100\unitlength,keepaspectratio=true]{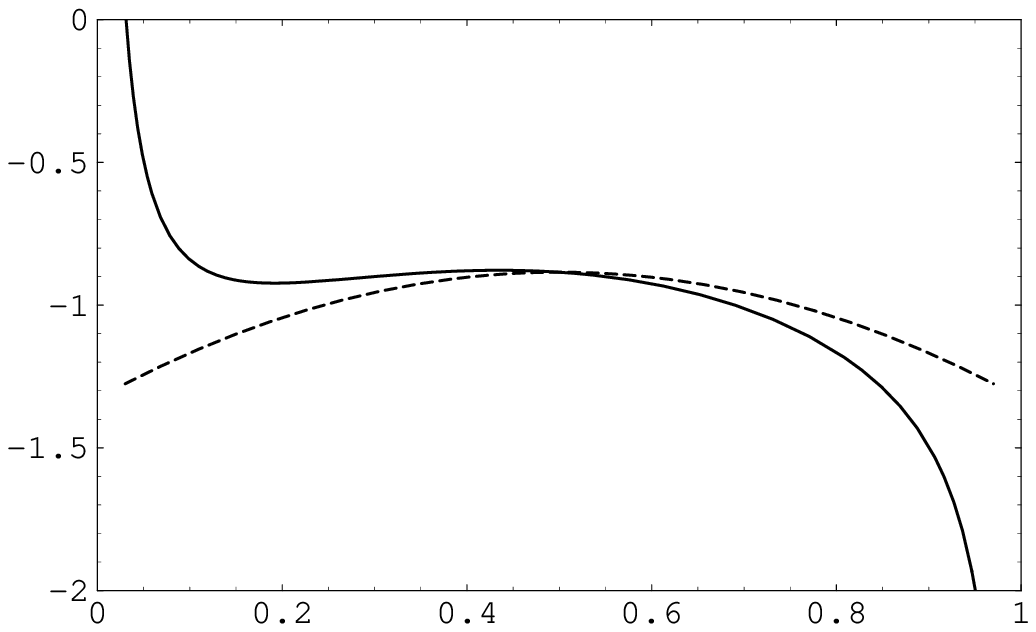}}
\end{picture}
\caption{The dependence of $\sigma_0^{-1} d\sigma_C/dx$ on $x$, see
(\ref{eq:spectr}), for $Z=82, \ \omega=50MeV$. Dashed curve: leading
approximation; solid curve: first correction is taken into account. }
\label{fig1}
\end{figure}

Using the results of \cite{LMS2000}, we obtain from (\ref{eq:se})
\begin{eqnarray}\label{eq:spectr}
&&\frac{d\sigma_C^{(0)}}{dx}+\frac{d\sigma_C^{(1)}}{dx}=-{4\sigma_0}\biggl[
\left(1-\frac43 x(1-x) \right)
f(Z\alpha)\nonumber\\
&& -\frac{\pi^3(1-2x)m}{8x(1-x)\omega} \left(1-\frac32 x(1-x)\right)
\,\mbox{Re}\, g(Z\alpha) \biggr],\nonumber\\
&&f(Z\alpha)= \mbox{Re}\psi(1+iZ\alpha)+C\,,\, g(Z\alpha)=
Z\alpha\,\frac{\Gamma(1-iZ\alpha)\Gamma(1/2 +i
Z\alpha)}{\Gamma(1+iZ\alpha)\Gamma(1/2 -i
Z\alpha)}\,,\nonumber\\
&&x=\varepsilon_-/\omega\quad ,\quad \sigma_0= \alpha(Z\alpha)^2/m^2\, ,
\end{eqnarray}
where $\psi(t)=d \ln \Gamma(t)/dt$, $C=0.577...$ is the Euler constant. In
(\ref{eq:spectr}), the term $\propto f(Z\alpha)$ corresponds to the leading
approximation $d\sigma_C^{(0)}/dx$ \citep{DBM1954}, the term
$\propto\mbox{Re}\, g(Z\alpha)$ is the first correction $d\sigma_C^{(1)}/dx$.
In contrast to the leading term, this correction is antisymmetric with respect
to the permutation $\varepsilon_+\leftrightarrow\varepsilon_-$ (or
$x\leftrightarrow 1-x$) and, therefore, does not contribute to the total cross
section. Besides, the correction is an odd function of $Z\alpha$ due to the
charge-parity conservation and the antisymmetry mentioned above. The
antisymmetric contribution enhances the production of electrons  at $x<1/2$ and
suppresses it at $x>1/2$. Evidently, the opposite situation occurs for
positrons. Qualitatively, such a behavior of the spectrum  takes place for any
$\omega$ being the most pronounced at low photon energy \citep{Overbo1968}. At
intermediate photon energies, the spectrum (\ref{eq:spectr}) essentially
differs from that given by the leading approximation. We illustrate this
statement in Fig.~\ref{fig1}, where $\sigma_0^{-1}d\sigma_C/dx$ with correction
(solid line) and without correction (dashed line) are plotted for $Z=82$ and
$\omega= 50\, MeV$.

Due to the antisymmetry of ${d\sigma_C^{(1)}}/{dx}$ at $\varepsilon_\pm\gg m$,
the term $\sigma_C^{(1)}$ in the total cross section may originate only from
the energy regions $\varepsilon_-\sim m$ and
$\varepsilon_+=\omega-\varepsilon_-\sim m$. The quasiclassical approximation
can not be used directly in these regions, and another approach is needed to
calculate the spectrum. We are going to do this elsewhere. However, for the
total cross section, it is possible to overcome this difficulty by means of
dispersion relations (see Section \ref{sec:CCT}).

As known \citep[see, e.g.,][]{BLP1980}, the spectrum of bremsstrahlung  can be
obtained from the spectrum of pair production. This can be performed by means
of the substitution $\varepsilon_+\to -\varepsilon$ , $\omega\to -\omega'$, and
$dx\to ydy$, where $y=\omega'/\varepsilon$, $\omega'$ is the energy of an
emitted photon, $\varepsilon$ is the initial electron energy. Using
(\ref{eq:spectr}), we obtain for the Coulomb corrections to the bremsstrahlung
spectrum
\begin{eqnarray}\label{eq:spectrphot}
y\frac{d\sigma_C^\gamma}{dy}&=&-{4\sigma_0}\biggl[
\left(y^2+\frac43(1-y) \right)
f(Z\alpha)\nonumber\\
&& -\frac{\pi^3(2-y)m}{8(1-y)\varepsilon} \left(y^2+\frac32
(1-y)\right) \,\mbox{Re}\, g(Z\alpha) \biggr]\, .
\end{eqnarray}

This formula describes bremsstrahlung from electrons. For positrons, it is
necessary to change the sign of $Z\alpha$ in (\ref{eq:spectrphot}). Our result
(\ref{eq:spectrphot}) coincides  with that obtained in \citep{BK1976} if the
obvious mistake in the latter is corrected by changing
$$\frac{1}{\gamma}\to \frac12\left(
\frac{m}{\varepsilon}+\frac{m}{\varepsilon-\omega'}\right)=
\frac{(2-y)m}{2(1-y)\varepsilon}\, $$ in Eq.(22) of \citep{BK1976}.

\section{Coulomb corrections to the total cross section}
\label{sec:CCT}

For $\omega\gg m$, we derive $\sigma_C^{(1)}$ using the relation
(\ref{eq:Opt}). In the leading approximation, the Coulomb corrections to the
cross section of pair production were obtained in \citep{DBM1954}. Using this
result and dispersion relations, the corresponding correction, $M_{DC}^{(0)}$,
to  the forward Delbr\"uck scattering amplitude $M_D$ were obtained in
\citep{Rohrlich1957}. They read:
\begin{eqnarray}\label{eq:MD1}
\sigma_C^{(0)}=-\frac{28}{9}\,\sigma_0\,f(Z\alpha)\, ,\quad
M_{DC}^{(0)}=-i\frac{28}{9}\omega\sigma_0 \,f(Z\alpha) \, ,
\end{eqnarray}
where $\sigma_0$ and  $f(Z\alpha)$ are defined in (\ref{eq:spectr}).

Using the results of \cite{LMS2000} for the quasiclassical Green functions, we
find within logarithmic accuracy
\begin{equation}\label{eq:delM}
\mbox{Re}\, M_{DC}^{(1)}=\,
\frac{\alpha(Z\alpha)^2\pi^3\,\mbox{Im}\,g(Z\alpha)}{m}\,\ln\frac{\omega}{m}
\,,
\end{equation}
the function $g(Z\alpha)$ is defined in (\ref{eq:spectr}). The logarithm
appears as a result of integration in (\ref{eq:Md}) over $\varepsilon$ in the
region where $\varepsilon\gg m$ and $\omega-\varepsilon\gg m$, thereby the
quasiclassical approximation is applicable. Unlike $\mbox{Re}\, M_{DC}^{(1)}$,
the quantity $\mbox{Im}\, M_{DC}^{(1)}$ is determined by the regions of
integration over $\varepsilon$, where $\varepsilon\sim m$ and
$\omega-\varepsilon\sim m$, and, therefore, the quasiclassical approximation is
invalid. Nevertheless, this quantity, which is related to $\sigma_C^{(1)}$
(\ref{eq:Opt}), can be obtained from the dispersion relation for $M_D$
\citep{Rohrlich1957}
\begin{eqnarray}\label{eq:disp}
\mbox{Re}M_D(\omega)&=&\frac{2}{\pi}\omega^2\, P\!\int_0^\infty
\frac{\mbox{Im}M_D(\omega^{\prime}) \,d\omega'} {\omega^{\prime}(\omega^{\prime
2}-\omega^2)}\,.
\end{eqnarray}

Using this relation, it can be easily checked that the high-energy asymptotics
(\ref{eq:delM}) unambiguously corresponds to the $\omega$-independent
high-energy asymptotics
\begin{equation}\label{eq:delM1}
 \mbox{Im}\, M_{DC}^{(1)}=
-\frac{\alpha(Z\alpha)^2\pi^4\,\mbox{Im}\,g(Z\alpha)}{2m} \, ,
\end{equation}

\begin{figure}[h]
\centering \setlength{\unitlength}{0.1cm}
\begin{picture}(105,80)
 \put(56,0){\makebox(0,0)[t]{$Z$}}
 \put(-6,35){\rotatebox[origin=c]{90}{$f_1$}}
 \put(0,0){\includegraphics[width=100\unitlength]{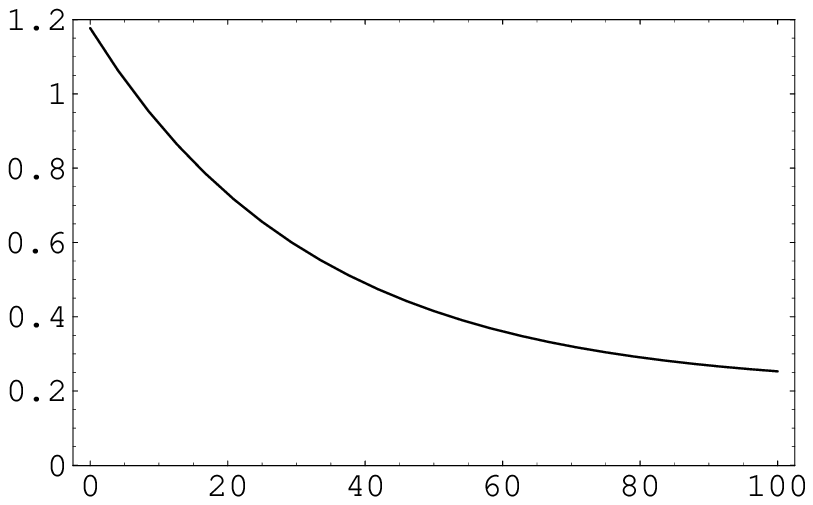}}
\end{picture}
\caption{The quantity $f_1$ as a function of $Z$}
 \label{fig2}\end{figure}

Substituting (\ref{eq:delM1}) into (\ref{eq:Opt}) and using $\sigma_{bf}$ from
\citep{MS1993} in the form
\begin{eqnarray}\label{eq:sigbf}
\sigma_{bf}&=&\,
4\pi\sigma_0\,(Z\alpha)^3\,f_1(Z\alpha)\,\frac{m}{\omega}\quad ,
\end{eqnarray}
we have for $\sigma_C^{(1)}$
\begin{eqnarray}\label{eq:delsigma} 
\sigma_C^{(1)}&=&-\sigma_0\,\left[\frac{\pi^4}{2}\, \mbox{Im}\,g(Z\alpha)
+4\pi(Z\alpha)^3\,f_1(Z\alpha)\right]\,\frac{m}{\omega}\quad .
\end{eqnarray}
The function $f_1(Z\alpha)$ is plotted in Fig.~\ref{fig2}.

\begin{figure}[h]
\centering \setlength{\unitlength}{0.1cm}
\begin{picture}(105,80)
 \put(56,0){\makebox(0,0)[t]{$Z$}}
 \put(-6,30){\rotatebox[origin=c]{90}{$(\omega/m)\delta\sigma_C/\sigma_C$}}
\put(0,0){\includegraphics[width=100\unitlength]{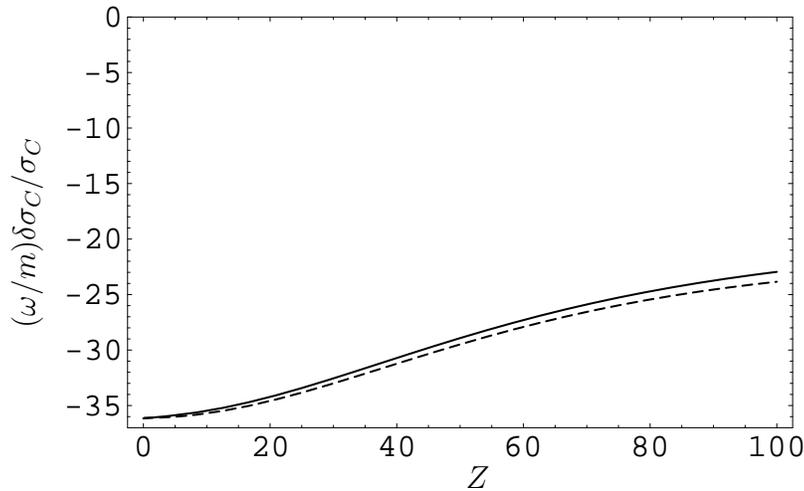}}
\end{picture}
\caption{The quantity $(\omega/m)\sigma_C^{(1)}/\sigma_C^{(0)}$ as a function
of $Z$(solid curve). Dashed curve corresponds to the same quantity without the
contribution of the bound-free pair production.}
 \label{fig2a}\end{figure}

The quantity $(\omega/m)\sigma_C^{(1)}/\sigma_C^{(0)}$ is shown in Fig.
\ref{fig2a} (solid curve). It is seen that this ratio is numerically large for
any $Z$. Therefore, the  term $\sigma_C^{(1)}$ gives a significant contribution
to $\sigma_C$ for intermediate photon energies. Dashed curve in Fig.
\ref{fig2a} gives the same ratio when $\sigma_{bf}$ in (\ref{eq:delsigma}) is
omitted. It is seen that the relative contribution of the term $\propto
f_1(Z\alpha)$ in (\ref{eq:delsigma}) is numerically small.

\section{Screening corrections}\label{sec:Scr}

In two previous Sections the cross section of $e^+e^-$ pair production has been
considered for a pure Coulomb field. The difference, $\delta V(r)$, between an
atomic potential and a Coulomb potential of a nucleus leads to the modification
of this cross section known as the effect of screening. In the Born
approximation, this effect was studied long ago \citep[see, e.g.,][]{JLS1950}.
Let us consider now $\sigma_C^{(scr)}$ characterizing the influence of
screening on the Coulomb corrections. Recollect that the Coulomb corrections
denote the higher-order terms of the perturbation theory with respect to the
atomic field. So far it was only known that the correction $\sigma_C^{(scr)}$
is not large \citep{DBM1954}. Here we consider this issue quantitatively. The
Coulomb corrections are determined by distances $r\sim 1/m$ where the
difference $\delta V(r)$ is small. In our calculation of $\sigma_C^{(scr)}$, we
retain the linear term of expansion in powers of $\delta V(r)$. We represent
$\delta V(r)$ as
\begin{equation}\label{eq:FF}
\delta V(r)=\int \frac{d\bm Q}{(2\pi)^3}\, e^{i\bm Q\bm r}\,F(Q)\frac{4\pi
Z\alpha}{Q^2}\, ,
\end{equation}
\begin{figure}[h]
\centering \setlength{\unitlength}{0.1cm}
\begin{picture}(105,80)
 \put(56,0){\makebox(0,0)[t]{$Z$}}
 \put(-6,30){\rotatebox[origin=c]{90}{$\sigma_C^{(scr)}/\sigma_C^{(0)}$}}
\put(0,0){\includegraphics[width=100\unitlength,keepaspectratio=true]{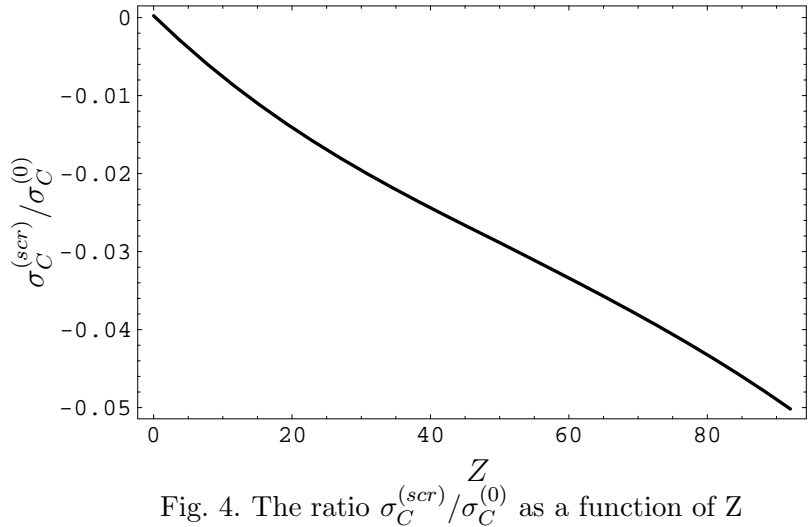}}
\end{picture}
\caption{The ratio $\sigma_C^{(scr)}/\sigma_C^{(0)}$ as a function of Z}
\label{fig3}\end{figure} where $F(Q)$ is the atomic electron form factor. Using
the results obtained in \citep{LM1995,KS2001,LMS2000} for arbitrary atomic
potential, we have for the correction to the spectrum
\begin{eqnarray}\label{eq:scrspect}
&&\frac{d\sigma_C^{(scr)}}{dx}=\frac{32}{3}\sigma_0m^2 \int_0^\infty
\frac{dQ}{Q^3}F(Q)\int_0^\infty\frac{d\tau}{\sinh
\tau}\left[\frac{\sin(2Z\alpha\tau)}{2Z\alpha}-\tau\right]\nonumber\\
&&\times\int_0^{2\pi}\frac{d\varphi}{2\pi}
\left[e^{\tau}R(\mu_+,\,a)-e^{-\tau}
 R(\mu_-,\,a)\right] \, ,\\
 &&R(\mu,\,a)=\frac{(\mu-1)}{4\mu^{2}}\Biggl\{
 \frac{1}{2\sqrt{\mu}}\left[18-6\mu+a(\mu^2+2\mu-3)\right]\,
 \ln\left[\frac{\sqrt{\mu}+1}{\sqrt{\mu}-1}\right]\nonumber\\
 &&-18-a(\mu-3)\Biggr\}\nonumber\\
 && \mu_\pm=
 1+\frac{8m^2\e^{\pm\tau}\sinh^2\tau}{Q^2(\cosh\tau+\cos\varphi)}\,
 ,\quad a=6x(1-x)\, .
\end{eqnarray}
Integrating over $x$, we obtain
\begin{eqnarray}\label{eq:scr}
&&\sigma_C^{(scr)}=\frac{32}{3}\sigma_0m^2 \int_0^\infty
\frac{dQ}{Q^3}F(Q)\int_0^\infty\frac{d\tau}{\sinh
\tau}\left[\frac{\sin(2Z\alpha\tau)}{2Z\alpha}-\tau\right]\nonumber\\
&&\times\int_0^{2\pi}\frac{d\varphi}{2\pi}
\left[e^{\tau}R(\mu_+,\,1)-e^{-\tau}
 R(\mu_-,\,1)\right] \, \, .
\end{eqnarray}
Similar to $\sigma_C^{(0)}$, this correction is $\omega$-independent. Shown in
Fig.~\ref{fig3} is the $Z$-dependence of the ratio
$\sigma_C^{(scr)}/\sigma_C^{(0)}$ calculated with the use of the form factors
taken from \citep{HO1979}. As seen from Fig.~\ref{fig3}, this ratio is
approximately described by the linear function $-5.4\cdot 10^{-4}\cdot Z$.

The corresponding correction to the bremsstrahlung spectrum is obtained from
(\ref{eq:scrspect}) by means of the same substitutions as in Section
\ref{sec:CCS}. So that the quantity $y^{-1}d\sigma_C^{\gamma(scr)}/dy$ is given
by the right-hand side of (\ref{eq:scrspect}) if we set $a=6(y-1)/y^2$.

\section{Comparison with experimental data
and estimation of $\sigma_C^{(2)}$}

The most detailed and accurate experimental data have been obtained just in the
region of intermediate photon energies. In this region, the first correction
$\sigma_C^{(1)}$, obtained above, becomes large, see Fig. \ref{fig2a}, and the
next term $\sigma_C^{(2)}$ in the expansion (\ref{eq:expansion}) may be
significant. Since it has not been calculated, we use for $\sigma_C^{(2)}$ the
following ansatz:
\begin{equation}\label{eq:ansatz}
\sigma_C^{(2)}=\sigma_0 \left[b\ln(\omega/2m)+c\,\right] \left(\frac
m\omega\right)^2\,,
\end{equation}
where $b$ and $c$ are some functions of $Z\alpha$. This form follows from the
arguments similar to those presented by \citet{DBM1954}.

\begin{figure}[h]
\centering \setlength{\unitlength}{0.1cm}
\begin{picture}(105,80)
 \put(56,0){\makebox(0,0)[t]{$\omega$ $(MeV)$}}
 \put(-4,30){\rotatebox[origin=c]{90}{$\Sigma$}}
\put(0,0){\includegraphics[width=100\unitlength,keepaspectratio=true]{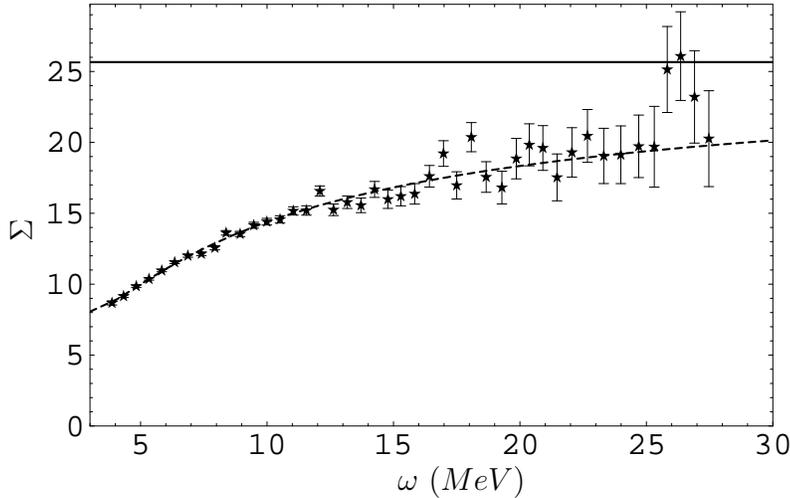}}
\end{picture}
\caption{The values of $\Sigma$ (\ref{eq:SigmaExp}) extracted from the
experimental data for $Bi$ together with the fit $a+(m/\omega)[b
\ln(\omega/2m)+c]$ (dashed curve) and the theoretical asymptotics
$a_{th}=({\omega}/m)\sigma_0^{-1}\sigma_C^{(1)}$ (solid line). Coefficients
$a$,$b$, and $c$ are given in the text.} \label{fig4}\end{figure}

Shown in Fig. \ref{fig4} is the quantity
\begin{equation}\label{eq:SigmaExp}
 \Sigma=\frac\omega
m \sigma_0^{-1}(\sigma_{coh}-\sigma_B-\sigma_C^{(0)}-\sigma_C^{(scr)})\,,
\end{equation}
where the experimental cross section $\sigma_{coh}$ for $Bi$ is taken from
\citep{SRL1980}, $\sigma_B$ is the Born cross section calculated with screening
taken into account, $\sigma_C^{(0)}$ and $\sigma_C^{(scr)}$ are given by
(\ref{eq:MD1}) and (\ref{eq:scr}), respectively. This quantity is fitted by the
formula $a+(m/\omega)[b\ln(\omega/{2m})+c]$ (dashed curve) which corresponds to
the sum $(\omega/ m) \sigma_0^{-1}(\sigma_C^{(1)}+\sigma_C^{(2)})$. The fitting
parameters obtained by the linear regression method are $a=25.79$,
$b=-98.92$,$c=2.43$. The quantity $a$ determines the asymptotics of the fit at
$\omega \to \infty$ and should be compared with our result
$a_{th}=(\omega/m)\sigma_0^{-1}\,\sigma_C^{(1)}=25.66$ (solid line). So the
value of $a_{th}$ is in a perfect agreement with that extracted from the
experimental data. From Fig. \ref{fig4}, we  conclude that the term
$\sigma_C^{(2)}$ gives a noticeable contribution to $\Sigma$.

Note that the coefficient $c$ is small as compared to $b$. The same situation
($a\approx a_{th}$, $|b|\gg |c|$) takes place at fitting of experimental data
for $Ta$ \citep{SRL1980}. So, we set $c=0$ in Eq. (\ref{eq:ansatz}) and use $b$
as the only fitting parameter. Fitting the data for $Bi$ with such ansatz, we
obtain
$b=-96.63$. A new curve practically coincides with that shown in Fig.
\ref{fig4}.

\begin{figure}[h]
\centering \setlength{\unitlength}{0.1cm}
\begin{picture}(105,80)
 \put(56,0){\makebox(0,0)[t]{$\omega$ $(MeV)$}}
 \put(-4,30){\rotatebox[origin=c]{90}{$\Sigma$}}
\put(0,0){\includegraphics[width=100\unitlength,keepaspectratio=true]{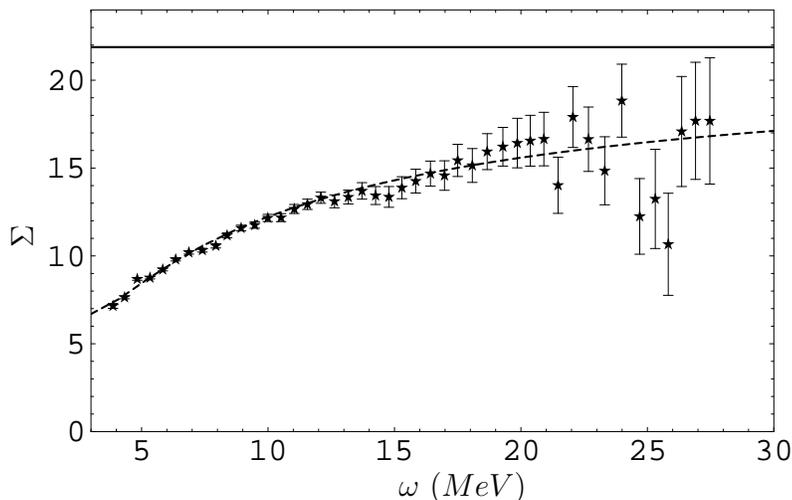}}
\end{picture}
\caption{The values of $\Sigma$ (\ref{eq:SigmaExp}) extracted from the
experimental data for $Ta$ together with the fit $a_{th}+(m/\omega)b
\ln(\omega/2m)$ (dashed curve), $a_{th}=21.88$,
$b=-82.82$. The solid line represents the asymptotics $\Sigma=a_{th}$.}
 \label{fig5}\end{figure}

\begin{figure}[h]
\centering \setlength{\unitlength}{0.1cm}
\begin{picture}(105,80)
 \put(56,0){\makebox(0,0)[t]{$\omega$ $(MeV)$}}
 \put(-4,30){\rotatebox[origin=c]{90}{$\Sigma$}}
\put(0,0){\includegraphics[width=100\unitlength,keepaspectratio=true]{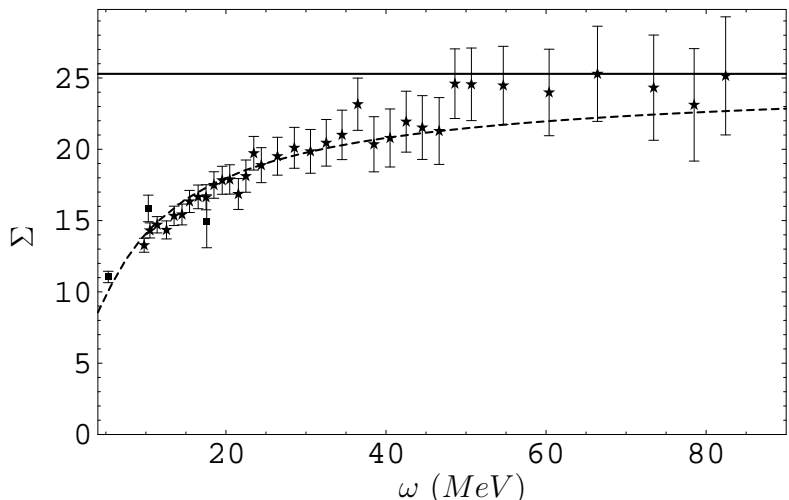}}
\end{picture}
\caption{The values of $\Sigma$ (\ref{eq:SigmaExp}) extracted from the
experimental data for $Pb$ obtained by \citeauthor{RSW1952} (squares) and
\citeauthor{SRL1980} (stars). The dashed curve corresponds to the fit
$a_{th}+(m/\omega)b \ln(\omega/2m)$, $a_{th}=25.29$, $b=-95.95$. The solid line
represents the asymptotics $\Sigma=a_{th}$.} \label{fig6}\end{figure}

The  values of $\Sigma$ extracted from the experimental data for $Ta$
\citep{SRL1980} and $Pb$ \citep{RSW1952,GH1978} are shown in Fig. \ref{fig5}
and Fig. \ref{fig6}, respectively. Dashed curves represent the results of the
fit, solid lines correspond to the asymptotics $\Sigma=a_{th}$.

Within the accuracy of the fitting procedure, we obtain $b = -3.78\cdot a_{th}$
for all three elements considered. This fact hints at the same $Z$-dependence
of the coefficients $a_{th}$ and $b$, at least for heavy atoms.

In the paper \citep{DBM1954} it was mentioned that all available experimental
data on $Pb$ above $5MeV$ are well represented by the formula
$\sigma_{coh}=\sigma_B+\sigma_C+11.8\cdot(m/\omega)\sigma_0$. Poor knowledge of
the nuclear photoabsorption cross section  at that time can not explain the
large difference between the coefficient $11.8$ and our result $a_{th}=25.29$.
The main source of this difference is the neglecting of the term
$\sigma_C^{(2)}$. It is evident from Fig. \ref{fig6} that an attempt to fit
$\Sigma$ in the region below $18\,MeV$ by the constant only, leads immediately
to the appreciably smaller coefficient $a$ than our value $a_{th}$.

\begin{figure}[h]
\centering \setlength{\unitlength}{0.1cm}
\begin{picture}(105,80)
 \put(56,0){\makebox(0,0)[t]{$\omega$ $(MeV)$}}
 \put(-4,30){\rotatebox[origin=c]{90}{$S$}}
\put(0,0){\includegraphics[width=100\unitlength,keepaspectratio=true]{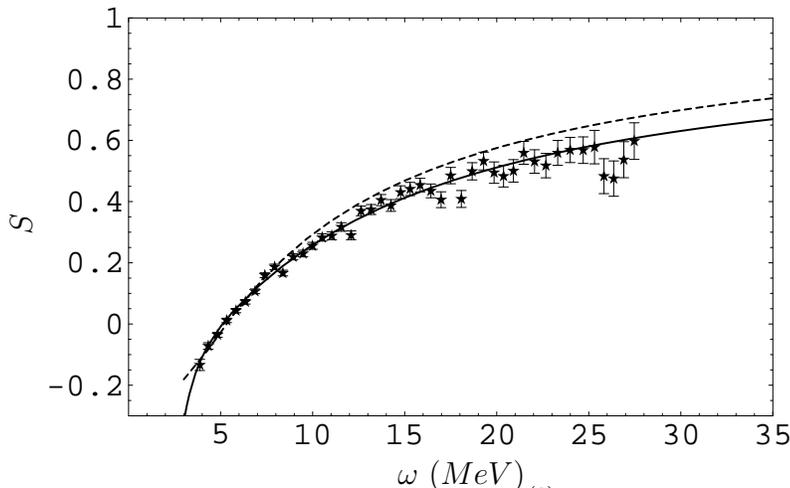}}
\end{picture}
\caption{The $\omega$-dependence of $S=(\sigma_{coh}-\sigma_B)/\sigma_C^{(0)}$
for $Bi$. Solid curve: our result; dashed curve: the result of
\citet{Overbo1977}; experimental data from \citep{SRL1980}.}
\label{fig7}\end{figure}

\begin{figure}[h]
\centering \setlength{\unitlength}{0.1cm}
\begin{picture}(105,80)
 \put(56,0){\makebox(0,0)[t]{$\omega$ $(MeV)$}}
 \put(-4,30){\rotatebox[origin=c]{90}{$S$}}
\put(0,0){\includegraphics[width=100\unitlength,keepaspectratio=true]{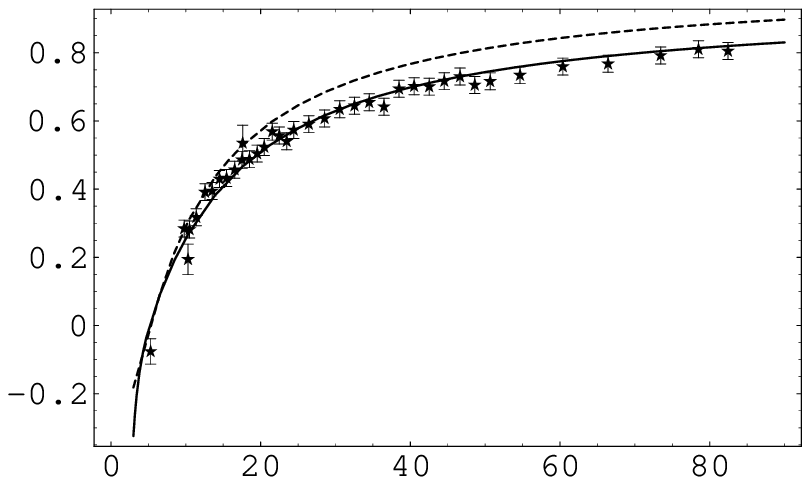}}
\end{picture}
\caption{Same as Fig. \ref{fig7} but for $Pb$; experimental data from
\citep{RSW1952,GH1978}.} \label{fig8}\end{figure}

It is interesting to compare our predictions for the Coulomb corrections to the
total cross section with the results of \citet{Overbo1977}. Shown in Figs.
\ref{fig7},\ref{fig8} is the ratio $S=(\sigma_{coh}-\sigma_B)/\sigma_C^{(0)}$,
which is the Coulomb corrections in units of $\sigma_C^{(0)}$, (\ref{eq:MD1}).
Our results are represented by solid curves, those of \citeauthor{Overbo1977}
are shown as dashed curves. The values of $S$ extracted from the experimental
data are also shown. The results for $Bi$ are plotted in Fig. \ref{fig7} with
the experimental data taken from  \citep{SRL1980}.  The results for $Pb$ are
plotted in Fig. \ref{fig8} with the experimental data taken from
\citep{RSW1952,GH1978}. It is seen that the difference between our results and
those of \citeauthor{Overbo1977} is small at relatively low energies and
becomes noticeable as $\omega$ increases. According to our results, this
difference tends to a constant $\sigma_C^{(scr)}/\sigma_C^{(0)}$ at $\omega\to
\infty$. As a whole, the experimental data are in a better agreement with our
results than with those of \citeauthor{Overbo1977}.

\section{Conclusion}

For the $e^+e^-$ photoproduction, we have calculated the leading correction
(\ref{eq:spectr}) to the electron spectrum in the region $\varepsilon_\pm\gg
m$. This contribution noticeably modifies the spectrum at intermediate photon
energy. It turns out that the correction is antisymmetric with respect to the
permutation $\varepsilon_+\leftrightarrow \varepsilon_-$ and hence does not
contribute to the total cross section. The leading correction to the total
cross section, $\sigma_C^{(1)}$, originates from two regions,
$\varepsilon_+\sim m$ and $\varepsilon_-\sim m$. We have obtained
$\sigma_C^{(1)}$ (\ref{eq:delsigma}) using dispersion relations. In contrast to
the form of the fit suggested by \citet{Overbo1977}, the quantity
$\sigma_C^{(1)}$ does not contain any powers of $\ln (\omega/ m)$. We have also
performed the quantitative investigation of the influence of screening on the
Coulomb corrections (\ref{eq:scrspect}),(\ref{eq:scr}). It is important that
$\sigma_C^{(scr)}$ does not vanish in the high-energy limit. We have suggested
a form for the next-to-leading correction, $\sigma_C^{(2)}$, to the total cross
section. Altogether, the corrections found allow one to represent well the
available experimental data.

Starting with the results obtained for  the  $e^+e^-$ photoproduction spectrum,
we have obtained the corresponding correction to the bremsstrahlung spectrum as
well.

\section*{Acknowledgement}
We are indebted to J.H.~Hubbell for his continuing interest to this work. This
work was supported in part by RFBR Grants 01-02-16926 and 03-02-16510.

\end{document}